# A new regime of nanoscale thermal transport: collective diffusion counteracts dissipation inefficiency

Kathleen M. Hoogeboom-Pot[1], Jorge N. Hernandez-Charpak[1], Erik H. Anderson[2], Xiaokun Gu[3], Ronggui Yang[3], Margaret M. Murnane[1], Henry C. Kapteyn[1] and Damiano Nardi[1,*]

[1]JILA and Department of Physics, University of Colorado, Boulder, CO 80309, USA
[2]Center for X-Ray Optics, Lawrence Berkeley National Lab, Berkeley, CA 94720, USA
[3]Department of Mechanical Engineering, University of Colorado, Boulder, CO 80309, USA

[*]*E-mail:* damiano.nardi@jila.colorado.edu

**Abstract**

Understanding thermal transport from nanoscale heat sources is important for a fundamental description of energy flow in materials, as well as for many technological applications including thermal management in nanoelectronics, thermoelectric devices, nano-enhanced photovoltaics and nanoparticle-mediated thermal therapies. Thermal transport at the nanoscale is fundamentally different from that at the macroscale and is determined by the distribution of carrier mean free paths in a material, the length scales of the heat sources, and the distance over which heat is transported. Past work has shown that Fourier's law for heat conduction dramatically over-predicts the rate of heat dissipation from heat sources with dimensions smaller than the mean free path of the dominant heat-carrying phonons. In this work, we uncover a new regime of nanoscale thermal transport that dominates when the separation between nanoscale heat sources is small compared with the dominant phonon mean free paths. Surprisingly, the interplay between neighboring heat sources can facilitate efficient, diffusive-like heat dissipation, even from the smallest nanoscale heat sources. This finding suggests that thermal management in nanoscale systems including integrated circuits might not be as challenging as projected. Finally, we demonstrate a unique and new capability to extract mean free path distributions of phonons in materials, allowing the first experimental validation of differential conductivity predictions from first-principles calculations.

**Introduction**

Critical applications including thermoelectrics for energy harvesting, nanoparticle-mediated thermal therapy, nano-enhanced photovoltaics, and thermal management in integrated circuits require a fundamental understanding of energy flow at the nanoscale. Recent work has shown the rate of heat dissipation from a heat source is reduced significantly below that predicted by Fourier's law for diffusive heat transfer when the characteristic dimension of the heat source is smaller than the mean free path of the dominant heat carriers (phonons in dielectric and semiconductor materials)[1-6]. However, a complete fundamental description of nanoscale thermal transport is still elusive, and current theoretical efforts are limited by a lack of experimental validation.

Diffusive heat transfer requires many collisions between heat carriers to establish a local thermal equilibrium and a continuous temperature gradient along which energy dissipates. However, when the dimension of a heat source is smaller than the phonon mean free path (MFP), the diffusion equation is intrinsically invalid as phonons move ballistically without collisions, and the rate of nanoscale heat dissipation is significantly lower than the diffusive prediction. Furthermore, heat-carrying phonons in real materials have a wide distribution of MFPs, from several nanometers to hundreds of microns. For a given nanoscale heat source size, phonons with MFPs shorter than the hot spot dimension remain fully diffusive and contribute to efficient heat dissipation and a high thermal conductivity (or equivalently, a low thermal resistivity). In contrast, phonons with long MFPs travel ballistically far from the heat source before scattering, with an effective thermal resistivity far larger than the diffusive prediction. Phonons with intermediate MFPs fall in between; here heat transfer is quasi-ballistic with varying degrees of reduced contributions to the conduction of heat away from the nanoscale source. Most work to date explored the reduction in heat transfer from functionally isolated micro- and nanoscale heat sources[1-5]. Indeed, characterizing heat transfer from micro/nanostructures with varying size can be used to experimentally measure cumulative phonon mean free path spectra of materials[2,7,8], with the proof-of-principle demonstrated for long-wavelength (> 1 μm) MFP phonons in silicon[3].

In this work, we show through experiment and theory that the size of the heat source is not the only important scale that determines nanoscale heat dissipation. We identify a new regime of thermal transport that occurs when the separation between nanoscale heat sources is smaller than the average phonon mean free path. Surprisingly, the interaction of phonons from neighboring heat sources can counteract the reduction due to ballistic effects in nanoscale heat dissipation from individual sources, to such an extent that this collective behavior increases heat transfer to near the diffusive limit. Most importantly, the appearance of this new 'collectively-diffusive' regime mitigates scaling problems for thermal management in nanoelectronics, which may not be as serious as projected[2,9,10]. Finally, we use this new phenomenon to extract the contribution to thermal transport from specific regions of the phonon MFP spectrum, opening up a new approach for thermal transport metrology and mean free path spectroscopy. This is because by varying both nanostructure size and separation, an effective phonon filter is introduced that suppresses specific MFP contributions to thermal conductivity. We compare our extracted phonon mean free path spectra with predictions from first-principles calculations and find excellent agreement between experiment and theory. Looking forward, we have a unique new capability for characterizing phonon transport in materials, including novel, complex nanostructured materials and metamaterials, where predictions do not yet exist.

Figure 1 illustrates the differences between the three regimes of heat transport from nanoscale heat sources – purely diffusive, quasi-ballistic and collectively-diffusive. Quasi-ballistic transport (Fig. 1b) dominates when the size of isolated nanoscale heat sources is smaller than dominant phonon MFPs. In the new collectively-diffusive regime we uncovered (Fig. 1c), the separation between heat sources is small enough that long-MFP phonons, whose contribution to heat dissipation would normally be limited by the small size of nano-heat sources, can once again play a significant role. Although these phonons travel ballistically away from each individual heat source, they can scatter with phonons originating from a neighboring heat source, thus creating an effectively larger heat source size. In the limiting case, the spacing between heat sources vanishes and this regime approaches heat dissipation from a uniformly heated layer.

**Experiment**

In our experiment, arrays of nickel nanowires were fabricated by e-beam lithography and lift-off techniques to form periodic gratings on the surface of sapphire and silicon substrates. The nanowire linewidths $L$ range from 750 nm down to 30 nm, with period $P = 4L$ and a rectangular profile height of ≈ 13.5 nm. The use of nano-patterned structures rather than optical absorption allows us to explore heat sources much smaller than the diffraction limit of visible light. The metallic nanowires are heated by a 25 fs pump pulse centered at a wavelength of 800 nm. The sapphire substrate is transparent at this wavelength, while the silicon substrate has such a long absorption depth that any small, uniform heating of the substrate can be neglected. Laser excitation thus creates an array of nanoscale hot spots (lines) on the surface of a cold substrate. Because all nanostructures are fabricated on the same substrate at the same time, the intrinsic thermal boundary resistivity at the interface between the metallic line and the substrate will be constant across all samples: any variation in efficiency of heat dissipation as the hot spot size or spacing is varied can thus be attributed to different regimes of thermal transport.

The laser-induced thermal expansion and subsequent cooling of the nano-gratings is probed using coherent extreme ultraviolet (EUV) light centered at a wavelength of 29 nm, created by high harmonic up-conversion of an 800 nm Ti:sapphire laser[11]. The time delay between the EUV probe pulse and the laser pump pulse is adjusted using a mechanical delay stage between -400 ps and +8000 ps, with step size as small as 1 ps. As the EUV light diffracts from the periodic array of Ni nanowires, expansion and cooling of the nano-gratings changes the diffraction efficiency, and this signal is recorded by a CCD camera as a function of delay time between pump and probe pulses. Examples of this dynamic signal are shown in Fig. 2a. (Note that more details of the experimental setup are described elsewhere[12,13] and in Supplementary Section S1). Because the reflectivities of these materials do not change with temperature at EUV wavelengths[14], the change in the diffraction signal can be uniquely attributed to physical deformations in the surface profile. Thus, the data of Fig. 2a can be used to directly extract the average thermal expansion and relaxation of each individual nanowire induced by laser heating and subsequent heat dissipation into the substrate, in addition to the surface deformations caused by acoustic waves launched by the initial impulsive expansion[12,13,15].

**Theory**

To understand the different regimes of thermal transport illustrated in Fig. 1, we consider three models: 1) the model described in our previous work that assumes isolated heat sources[1]; 2) an analytical model we develop here to account for interactions of phonons originating from neighboring heat sources using a gray, single-phonon-MFP approximation; and 3) a more advanced interacting model that includes a distribution of phonon MFPs. As discussed in detail below, this interacting multi-MFP model allows us to extract MFP-dependent contributions to thermal conductivity for MFPs as short as 14 nm for the first time, providing data that can be directly compared with predictions from first-principles density functional theory.

To quantify the deviations from diffusive heat transport, we first build upon methods similar to those described by Siemens et al.[1], but including more comprehensive finite element physical modeling[16] to improve data reduction accuracy. We model our system using diffusive heat conduction theory, while allowing the effective thermal boundary resistivity (which sets the temperature discontinuity across the boundary between the nickel nanowires and the substrate) to vary as a function of linewidth to account for non-diffusive effects. We use accurate sample dimensions (height, linewidth and period) characterized by atomic force microscopy (see Supplementary Section S2). Fresnel optical propagation is then used to calculate the diffraction signal from the simulated surface deformations. The effective boundary resistivity, $r_{eff}$, that provides the best fit to the experimental data of Fig. 2a represents the sum of the constant intrinsic thermal boundary resistivity that originates from the material difference between nickel and substrate, $r_{TBR}$, and corrections, $r_{Corr}$, due to non-diffusive size effects when either $L$ or $P$ is smaller than MFPs. By assigning the non-diffusive contribution to the thermal boundary resistivity rather than to changes in the substrate conductivity, we maintain a simple modeling geometry and avoid the need to assume a particular region of the substrate in which a conductivity change should apply.

The effective resistivity results are plotted in Fig. 2b. For large linewidths on both sapphire and silicon substrates, the effective resistivity converges toward a constant value – the intrinsic thermal boundary resistivity. As the linewidth approaches the dominant phonon MFPs in the substrate, the effective resistivity rises as thermal transport becomes quasi-ballistic and the contribution to heat dissipation of long-MFP phonon modes is suppressed[1,3,7]. This behavior was successfully described in past work using a simple gray model for sapphire and fused silica, which assumes a single phonon MFP to loosely describe a weighted average of the MFPs from all the phonon modes contributing to thermal transport in a given material. According to this model, a ballistic correction term proportional to $\Lambda_{gray}/(L/2)$ can be added to the intrinsic thermal boundary resistivity[1,17]; this prediction is plotted in red in Fig. 2b.

However, as the linewidth (and period) shrinks further, Fig. 2b shows that the effective resistivity starts to decrease rather than continuing to increase. The constant grating duty cycle for our series of samples means that the smallest-linewidth nanowires are also those with the smallest separation between neighboring heat sources. Thus, for small linewidths the separation becomes comparable to dominant phonon MFPs. For silicon, this peak in $r_{eff}$ is shifted toward longer linewidths/periods compared to sapphire because the phonon MFP distribution in silicon is also shifted toward longer MFPs, i.e. silicon has a longer average MFP than sapphire[1,6]. As illustrated in Fig. 1c, in this new collectively-diffusive regime, longer-MFP phonons from neighboring heat sources interact with each other as they would if they originated from a single, large heat source, leading to diffusive-like heat dissipation and decreasing the effective resistivity. The quasi-ballistic model for isolated heat sources clearly fails to capture this experimental observation, and a new model for $r_{Corr}$ is required to account for the transition to this new collectively-diffusive regime.

We propose to use the concept of a notch filter in the MFP spectrum to describe the effects of grating linewidth and separation, shown schematically in Fig. 3. The notch filter suppresses the contribution of phonon modes with MFPs that fall between the linewidth $L$ and period $P$ of the nano-gratings. Thus, if the grating period (separation) remains large while the linewidth is decreased, one would expect the effective boundary resistivity to continue to rise, as shown in the red dashed lines of Fig. 2b. This is because the contributions of all phonon modes with MFPs longer than the linewidth $L$ are suppressed in the quasi-ballistic regime of isolated heat sources. On the other hand, if the grating period shrinks, long-MFP phonon modes start to contribute again since phonons originating from neighboring heat sources interact with each other as they would in a bulk system, so the effective boundary resistivity should recover toward the bulk value, as seen experimentally in Fig. 2b.

To build an analytical expression for $r_{Corr}$ based on this idea, we use the concept of a phonon conductivity suppression function, $S(L, P, \Lambda)$, similar to those described by others[7,18]. This suppression function is applied to a bulk differential conductivity spectrum versus phonon MFP, $k(\Lambda_i)$, to calculate an effective nanoscale conductivity $K_{nano}$:

$$K_{nano} = \sum_i k(\Lambda_i) \cdot S(L, P, \Lambda_i) \qquad (1)$$

We then relate $r_{Corr}$ to the change in conductivity represented by this suppression.

$$r_{Corr} = A \left( \frac{1}{K_{nano}} - \frac{1}{K_{bulk}} \right) \qquad (2)$$

where $A$ collects geometrical constants and $K_{bulk}$ is the bulk conductivity of the substrate, simply given by $\sum_i k(\Lambda_i)$. For a given phonon MFP $\Lambda_i$, $S$ must approach unity in the diffusive regime when both $L$ and $P$ are large and at the limit of uniform heating when $L = P$. For the limit of small, isolated heat sources when $L \to 0$ but $P$ is large, $S \to L/(2\Lambda)$ to reproduce the behavior of the previously published quasi-ballistic model[1,19]. Finally, the effects of $L$ and $P$ should be the same but opposite to each other so that $L$ suppresses phonon mode contributions in the same way as $P$ reactivates them. To capture these behaviors along with smooth transitions among regimes, the two effects are represented by a special case of the generic family of logistic functions, where the total suppression function is written as:

$$S = tanh\left(\frac{L}{2\Lambda}\right) + \left[1 - tanh\left(\frac{P}{2\Lambda}\right)\right]. \tag{3}$$

More details of this suppression function can be found in Supplementary Section S3.

To test this new model for $r_{Corr}$, we first assume the simple single-MFP (gray) model (where the MFP distribution is a delta function). The resulting predictions are shown in the blue curves in Fig. 2b. Specifically, $r_{Corr}$ in this case is given by:

$$r_{Corr,\ gray}(L,P) = \frac{A}{K_{bulk}}\left(\frac{1}{S(L,P,\Lambda_{gray})} - 1\right) \tag{4}$$

Fitting this interacting model to the $r_{eff}$ data for sapphire, we extract values for $r_{TBR}$ and $\Lambda_{gray}$ which are consistent with previous results[1]: $\Lambda_{gray,\ int} = 131 \pm 11$ nm, $r_{TBR,\ int} = 2.58 \pm 0.19 \times 10^{-9}$ m²K/W. This good agreement with the previous larger-linewidth data and the accurate fit for the full range of our data validate our interacting $r_{Corr}$ model as an improved method to account for nanoscale size effects in heat transport – for both quasi-ballistic and collectively-diffusive regimes. Interestingly, this single-MFP model provides a reasonable approximation for the entire range of heat transport in sapphire.

For silicon, the interacting $r_{Corr}$ follows the general shape of the data and yields $\Lambda_{gray,\ int} = 360 \pm 26$ nm, which is consistent with previously reported values[6]. However, the interacting gray-model approach, although more successful than the isolated model, fails to capture the additional small peak in effective resistivity that appears for very small linewidths and periods, around $L = 40$ nm. The failure of this approach is not surprising, since the single-MFP model is known to be a poor approximation for silicon with its broad distribution of phonon MFPs[7,18].

**Discussion of new regime and extension to phonon mode conductivity spectra**

Having developed and validated the new model to capture the transitions from diffusive, to quasi-ballistic, to collectively-diffusive regimes, we can now extend our calculations beyond the simple single-MFP model and use our data to extract the MFP-dependent contributions to heat conduction in different materials down to MFPs as small as 14 nm for the first time. Because linewidth and period set the location and width of the effective notch filter in the phonon MFP spectrum, each configuration uniquely samples the contribution to thermal conductivity of different MFP ranges of phonon modes with a resolution controlled primarily by the number of configurations tested. The larger the resistivity correction needed for a given nano-grating, the stronger the conductivity contribution of phonon modes which were suppressed.

In order to extract information about the differential conductivity spectrum, we use the full multi-MFP form of $r_{Corr}$, given by:

$$r_{Corr}(L,P) = A \left( \frac{1}{\sum_i k(\Lambda_i) S(L,P,\Lambda_i)} - \frac{1}{K_{bulk}} \right) \qquad (5)$$

We partition the full sum in bins according to the MFP-sensitivity of each grating configuration (see Supplementary Section S3). Then by fitting our set of $r_{eff}$ data as $r_{eff} = r_{TBR} + r_{Corr}$ we extract the average $k(\Lambda_i)$ which is associated with all modes $\Lambda_i$ within each bin, thus assessing the relative contributions to thermal conductivity of each region of the phonon MFP spectrum (plotted in Fig. 4). By limiting the number of bins to be no more than half our number of data points, we ensure a conservative, well-conditioned fit, although we note that changing the bin number does not substantially alter the trends we observe. As shown by the purple curves in Fig. 2b, for sapphire, this procedure matches the experimentally measured $r_{eff}$ as well as the gray model for interacting heat sources. For silicon, this more complete multiple-MFP interacting model is able to match our experimental measurements of heat transport over the entire bulk, quasi-ballistic and collectively-diffusive regimes, including the additional feature where the resistivity increases for ≈ 40 nm linewidth (160 nm period).

While the number of data points limits the number of regions we can reasonably consider, this approach still offers an important and unprecedented experimental method to characterize the differential thermal conductivity contributions of phonons with different MFPs and to benchmark predictions including those from first-principles density functional theory (DFT) calculations, shown for silicon in Figs. 3 and 4. In particular, our DFT calculation for silicon, validated with others in literature,[20] are in very good agreement with our experimental data across all MFP ranges we examine. However, some small discrepancies appear for phonon MFPs around ≈ 100 nm, where experimental verification was not possible before. Our data are also consistent with the observations by others which emphasize the significant contributions from long-MFP (> 1 μm) modes[2,3,6], but the limited number of data points from structures much larger than the average phonon MFP results in a relatively large uncertainty in this region. For the purpose of comparison in Fig. 4 we set the normalization for the experimental spectra in silicon by assuming the integrated conductivity up to 1 μm should match that calculated by DFT. For sapphire, both calculation and experimental data imply that phonons with MFPs shorter than 1 μm are responsible for > 95% of the thermal conductivity. This ability to experimentally extract a phonon MFP distribution offers a new useful method for validating existing first-principles predictions across the whole range of phonon MFPs significant for heat conduction, as well as the first access to such information for more complex materials where calculations have not yet been performed. Most importantly, we have a unique capability to experimentally extract MFP spectra and mode conductivity, even in more complex materials and meta-materials.

**Conclusion**

In summary, we uncover a new regime of nanoscale thermal transport that dominates when the separation between nanoscale heat sources is small compared with the dominant phonon mean free paths. We also present a new approach for characterizing the relative contributions of phonons with different mean free paths which participate strongly in heat conduction. In particular, we can probe the small-MFP region which has been previously inaccessible to experiment. This unique capability is important as the need for precise phonon MFP distributions in complex nanostructures becomes more pressing – for both fundamental understanding and to harness systems where modeling does not yet exist. With bright soft X-ray high harmonic sources now reaching wavelengths below 1 nm[21], this approach can be extended even further into the deep nano-regime. Finally, the new, efficient, collectively-diffusive regime of thermal transport that we observe for the first time can potentially mitigate projected problems for thermal management in nanoelectronics, where the power density is likely to increase as the individual nanostructures shrink in size[9,10]. At the same time it illuminates important design implications for nanostructured materials and devices for energy and biomedical applications.

**Acknowledgements**

The authors gratefully acknowledge support from the US Department of Energy Basic Energy Sciences, the Semiconductor Research Corporation and used facilities provided by the NSF Engineering Research Center for EUV Science and Technology and a NSSEFF award. K.H-P. acknowledges support from the National Science Foundation under Award No. DGE 1144083. X.G. and R.Y. acknowledge the NSF CAREER award and AFOSR support.

**Author Contributions**

K.H.-P., J.H.-C., M.M., H.K. and D.N. planned the experiment. The samples were designed by K.H.-P., J.H-.C., E.A., M.M., H.K. and D.N. and fabricated by E.A. at LBNL. Experiment, simulations and data analysis were performed at JILA by K.H.-P., J.H.-C. and D.N. Density functional theory calculations were conducted by X.G. and R.Y. All authors discussed the results and contributed to the manuscript preparation.

**Additional Information**

The authors declare no competing financial interests. Correspondence and requests for materials should be addressed to D.N.

**Figures**

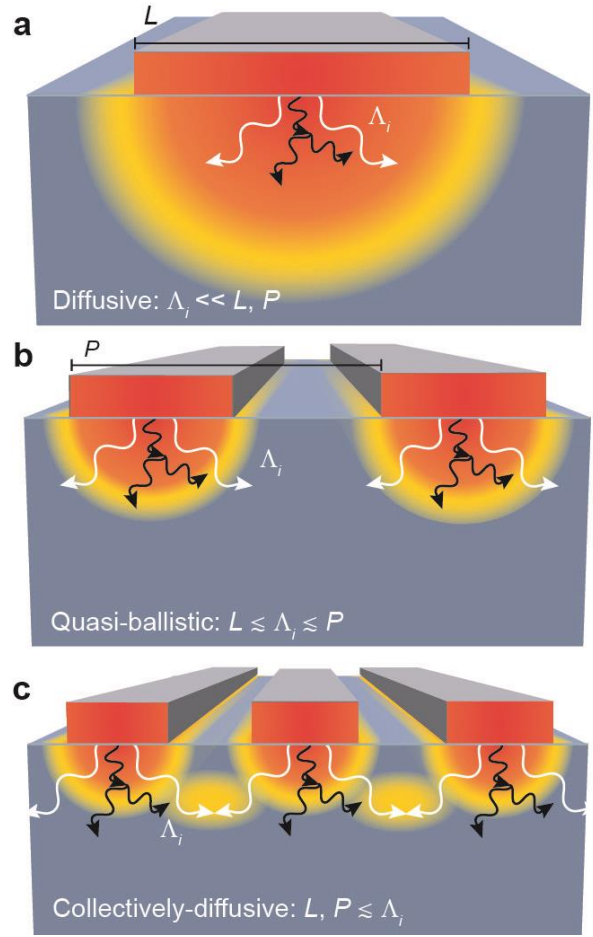

**Figure 1 | Three regimes in thermal transport.** Nanoscale heat transport is determined by the interplay between three length scales: the size of the heat sources $L$, the spacing of the heat sources $P$, and the MFPs $\Lambda_i$ of heat-carrying phonons. Materials support a broad distribution of MFPs, represented here by short (black) and long (white) MFP phonons. **a**, When all MFPs are smaller than $L$, heat dissipation is fully diffusive. **b,** As $L$ shrinks, long-MFP phonons travel ballistically, decreasing the rate of heat dissipation relative to diffusive predictions. Short-MFP phonons remain diffusive. **c**, When both $L$ and $P$ shrink, long-MFP phonons originating from neighboring heat sources interact as they would if they originated from a single, large heat source, enabling more efficient diffusive-like heat transfer from a larger effective heat source.

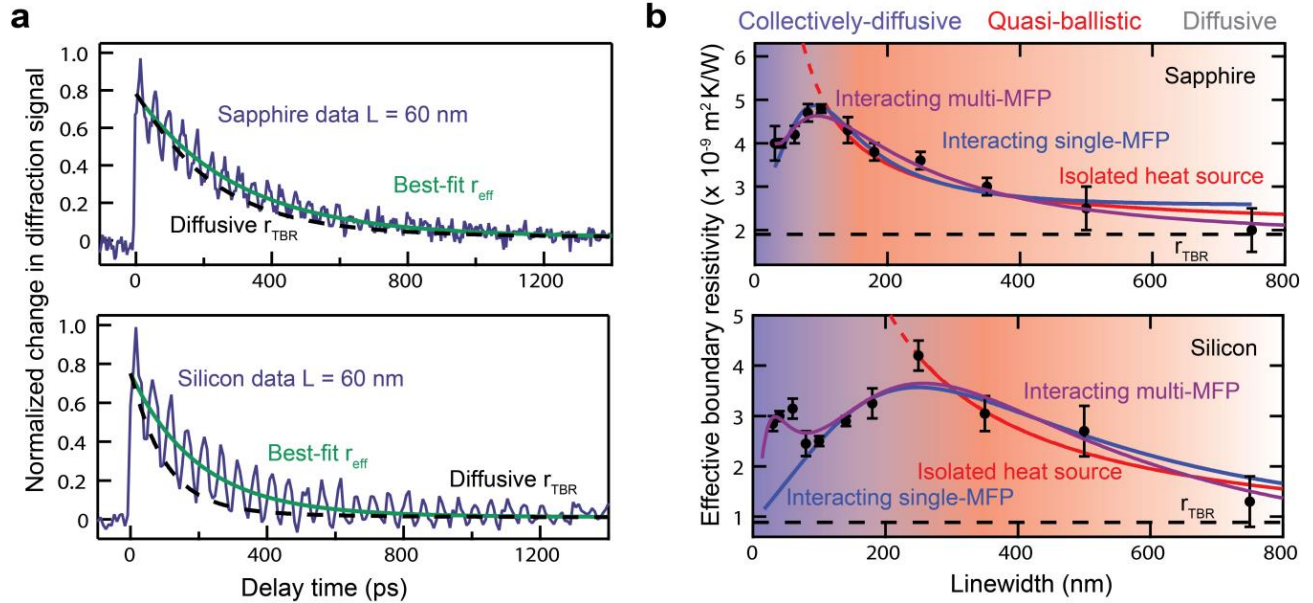

**Figure 2 | Effective thermal boundary resistivities extracted from dynamic EUV diffraction. a**, Dynamic diffraction from 60-nm-wide nickel lines on sapphire (top) and silicon (bottom) display a sudden rise due to impulsive thermal expansion following laser heating, a long decay due to thermal relaxation and oscillations due to surface acoustic waves. Dotted black lines plot the diffusive prediction, which significantly underestimates the thermal decay time. Green lines plot the decay using a best-fit to the effective thermal boundary resistivity. **b,** Extracted effective resistivities for each linewidth $L$ on both substrates increase with decreasing linewidth until the periods (equal to $4L$) are comparable to the average phonon MFP. For smaller periods (spacing), the effective resistivity decreases and approaches the diffusive limit (black dotted line). The error bars represent the standard deviation among multiple data sets for the same linewidth samples. Red lines: predictions for isolated heat sources based on the gray model. Blue lines: gray model including the onset of the collectively-diffusive regime. Purple lines: more complete model that includes contributions from multiple phonon MFPs.

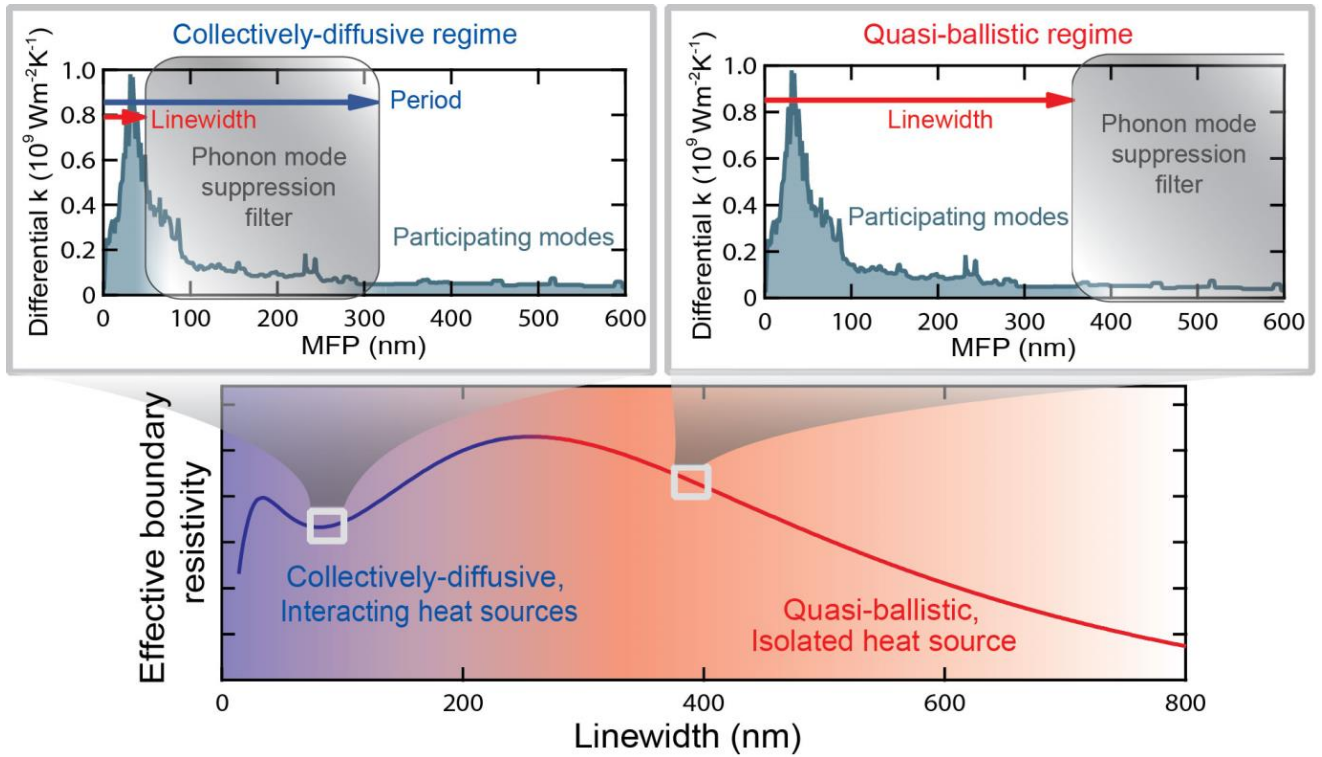

**Figure 3 | Linewidth and period define a suppression filter for phonon mean free path spectra.** The observed increase in effective thermal boundary resistivity for small linewidths *L* is due to the suppression of the contribution to thermal conductivity of phonon modes with MFP larger than *L*. Decreasing the period *P* can reactivate modes with MFP larger than *P*, decreasing the effective resistivity. In the limiting case of a uniformly heated layer, *P* approaches *L* and all phonon modes participate in thermal transport. We use as an example the smoothed differential conductivity distribution for silicon (top graphs, green line), calculated from first-principles density functional theory (see Supplementary Section S4).

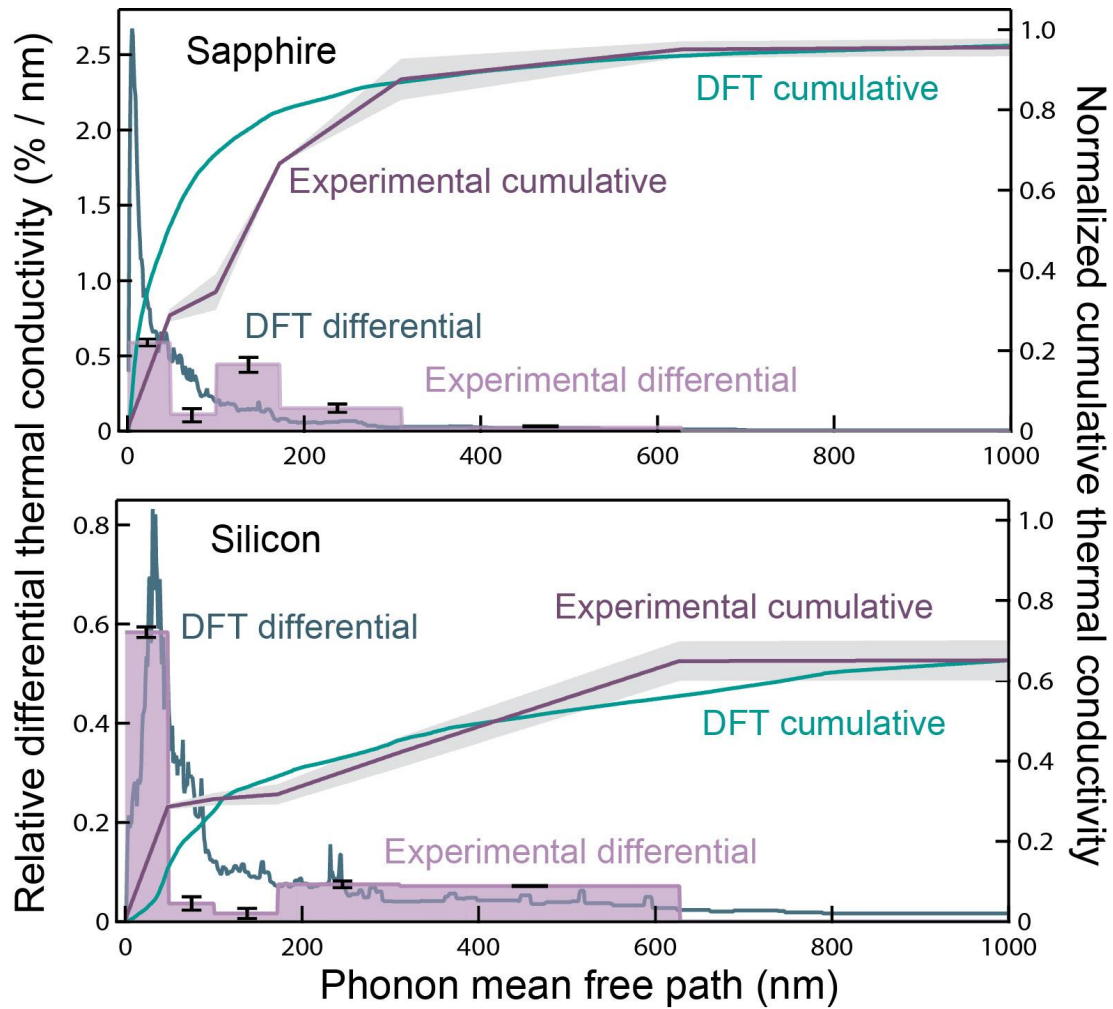

**Figure 4 | Extracting the differential and cumulative thermal conductivity.** By fitting $r_{eff}$ with multiple bins of phonon modes, the weights $k(\Lambda_i)$ assigned to those bins give the average relative contribution to the differential thermal conductivity (purple shading). Both differential (distributions) and cumulative (lines) conductivities are normalized to the total bulk conductivity. For sapphire (top panel), our data (purple) and first-principles DFT calculations (green) indicate there are no significant contributions from long-MFP phonons, so the cumulative curves approach unity at 1 μm. For silicon (bottom panel), our data are consistent with large contributions from longer MFPs.

# A new regime of nanoscale thermal transport: collective diffusion counteracts dissipation inefficiency

Kathleen M. Hoogeboom-Pot[1], Jorge N. Hernandez-Charpak[1], Erik H. Anderson[2], Xiaokun Gu[3], Ronggui Yang[3], Margaret M. Murnane[1], Henry C. Kapteyn[1] and Damiano Nardi[1,*]

[1]JILA and Department of Physics, University of Colorado, Boulder, CO 80309, USA
[2]Center for X-Ray Optics, Lawrence Berkeley National Lab, Berkeley, CA 94720, USA
[3]Department of Mechanical Engineering, University of Colorado, Boulder, CO 80309, USA
[*]E-mail: damiano.nardi@jila.colorado.edu

## Supplementary Information

### S1. Experimental setup and samples

For the time-resolved diffraction measurement, we used the pump-and-probe system described in our previous work[12,13]. The pump and probe pulses are derived from the same Ti:sapphire amplifier system (centered at a wavelength of 800 nm, with 4-6 kHz repetition rate, 1.5-2 mJ pulse energy and 25 fs pulse length) as shown in Supp. Fig. S1. The probe beam is focused into an argon-filled hollow waveguide to generate short-wavelength (centered at 29 nm) extreme ultraviolet (EUV) light via the nonlinear process of high-order harmonic generation[11]. The very short wavelength and interferometric diffraction measurement gives us high lateral and axial spatial resolution to follow the dynamics in nanostructured systems with sensitivity to surface displacements at the picometer scale. The EUV beam is focused onto the sample using a grazing-incidence toroidal mirror to a diameter of ≈ 100 μm (smaller than the 120 μm × 120 μm patterned area).

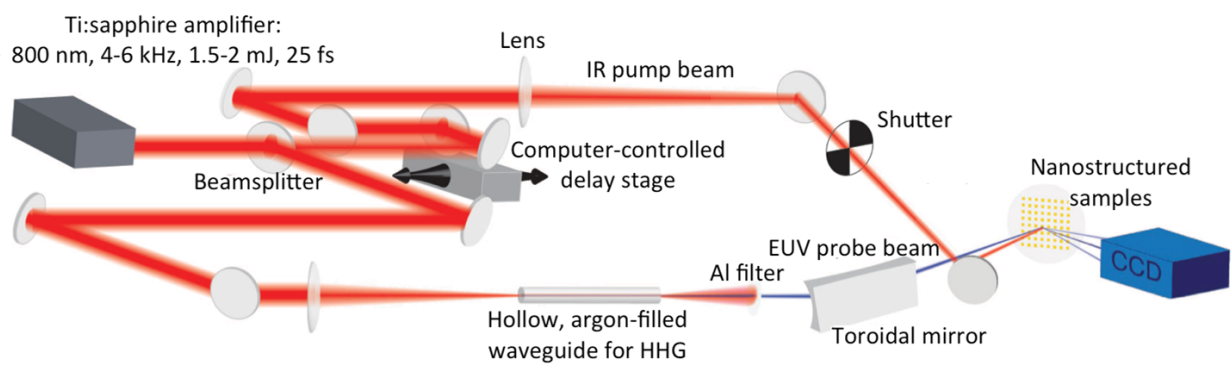

**Supplementary Figure S1 | Experimental setup.** Schematic illustration of the pump-probe diffraction measurement, adapted from Siemens et al.[1].

To ensure uniform heating of the nano-gratings, the pump beam is kept relatively large with a diameter of ≈ 400-500 μm and a fluence of ≤ 10 mJ/cm$^2$. The height of the nickel nanostructures is chosen to be less than or equal to the absorption depth of the infrared pump light for nearly uniform heating in the vertical direction as well. By testing the thermal decay dynamics with multiple pump fluences, we ensure that all our measurements are taken within the linear regime of heat transfer.

Because nickel is a metal, most of the heat is carried by electrons inside the nanostructures. Strong electron-phonon coupling ensures that the lattice has thermalized with the electrons within the first 10 ps – much faster than the timescale of thermal decay we are investigating. Because electron MFPs are much smaller than all the structures we explore, no significant nanoscale thermal conductivity effects should be expected in the nickel nanostructures.

To accurately characterize the sample (in terms of linewidth, height and period of the nano-grating) and design the unit cell in the finite-element simulations of the heat transport and thermal expansion, we imaged each sample using atomic force microscopy and scanning electron microscopy. We observed good uniformity across the nanostructures, as shown in the representative images in Supp. Fig. S2b-c.

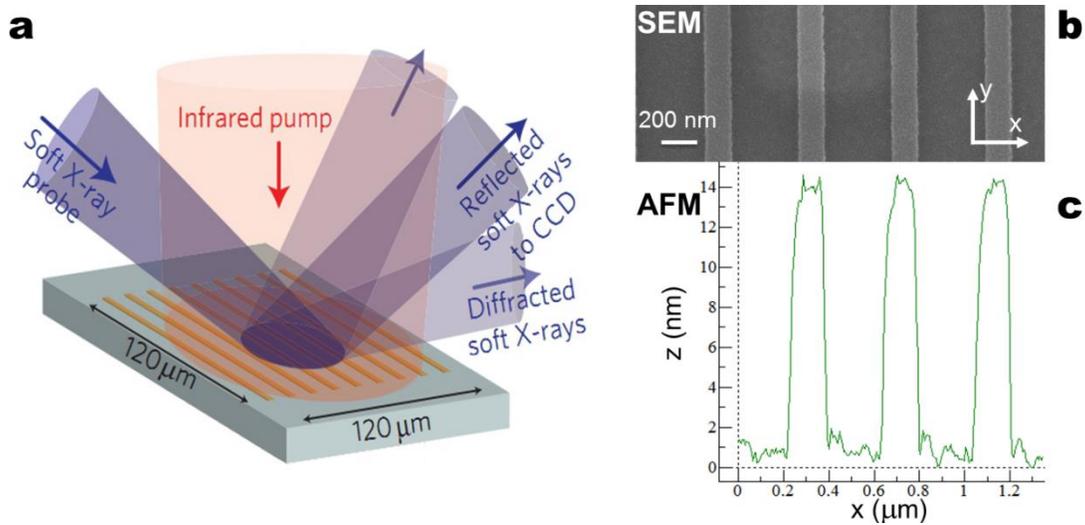

**Supplementary Figure S2 | Sample geometry. a**, Schematic illustration of the pump and probe beam illumination geometry at the sample. Each sample is characterized with (**b**) scanning electron microscopy and (**c**) atomic force microscopy to have accurate dimensions (linewidth, height and period of the nano-grating) as inputs for the finite-element simulations of the heat transport and thermal expansion. Shown here is one particular nickel-on-sapphire sample with $L$ = 100 nm (linewidth measured at the top of the nanowire).

## S2. Finite element multiphysics modeling for experimental data analysis

In order to quantitatively analyze the different regimes of nanoscale heat transport illustrated in Fig. 1 of the main text, we employ a similar approach as that described by Siemens et al.[1], but with the inclusion of more comprehensive finite element physical modeling[16] to reduce uncertainty in the comparison between experimental observations and the simulation output. We model our experimental observation using a two-dimensional simulation unit cell with the plane strain approximation to follow the full thermal expansion and cooling dynamics of the nickel nano-gratings on the sapphire and silicon substrates[15]. The nano-grating geometry allows for periodic boundary conditions for the temperature $T$ and the displacement $\boldsymbol{u}$ on the sides of the unit cell. The top boundaries are free to move but heat flux across them is set to zero (adiabatic boundary condition). The bottom boundary is fixed and the heat flux across it is also set to zero. Continuity in the displacement is enforced at the interface between the Ni structure and the substrate, effectively joining the two materials such that no slipping is allowed. The height of the substrate section is set to 10 μm to ensure no excess heat reaches the bottom boundary during the simulation time. An example of the top region of the unit cell geometry and mesh profile for the finite element simulations is illustrated in Supp. Fig. S3.

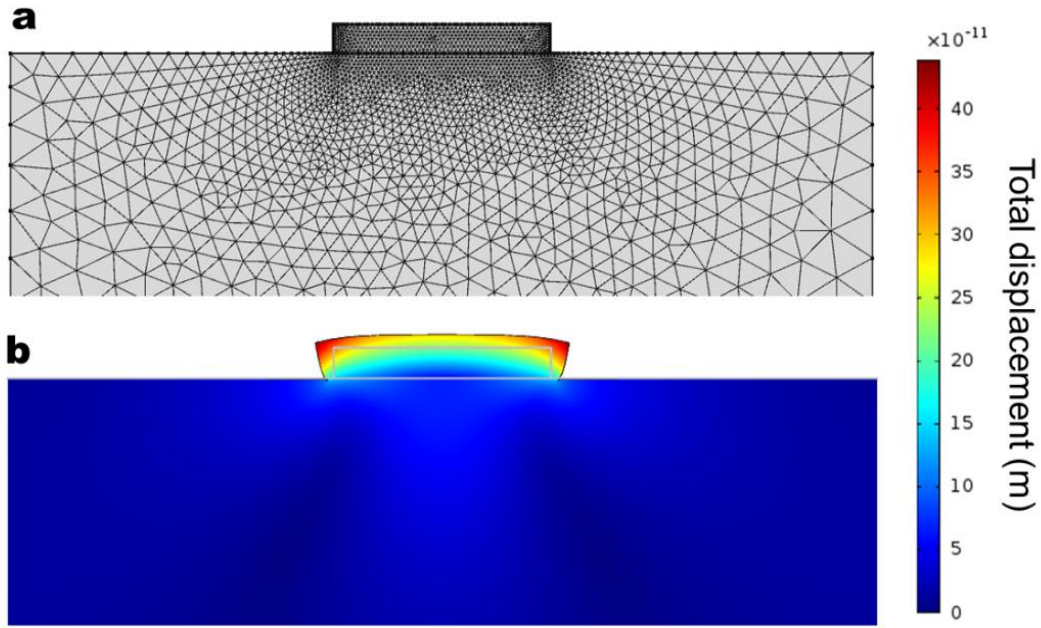

**Supplementary Figure S3 | Top section of finite element simulation cell. a**, The simulation mesh concentrates points inside the nanostructure and near the interface. **b**, The simulations output the time-dependent surface displacements following laser heating of the nano-grating (the surface deformation is here amplified for the purpose of visualization).

Unlike Siemens et al.[1], we incorporate the inertial terms in the initial thermal expansion and solve the coupled equations for the profiles of $T$ and $\boldsymbol{u}$ in the time domain[22]:

$$\nabla \cdot \left(\boldsymbol{c}: \nabla(\boldsymbol{u} - \alpha \Delta T)\right) = \rho \frac{\partial^2 \boldsymbol{u}}{\partial t^2} \quad \text{(S.1)}$$

$$\rho C_p \frac{\partial T}{\partial t} + \rho C_p \boldsymbol{u} \cdot \nabla T = \nabla \cdot (K_{bulk} \nabla T) + Q \quad \text{(S.2)}$$

where $\boldsymbol{c}$ is the elastic tensor, $\boldsymbol{u}$ is the displacement, $\rho$ is the density of the material, $\alpha$ is the linear coefficient of thermal expansion, $T$ is the temperature, $C_p$ is the specific heat of the material, $K_{bulk}$ is the bulk thermal conductivity and $Q$ is the heat source term accounting for the laser heating of the nano-grating, as described by Banfi et al.[23]. The more complete account of the physical dynamics in our sample allows for a more precise fit to our data and lowers our uncertainty, particular for small-linewidth samples.

The effective thermal boundary resistivity $r_{eff}$, which sets the temperature discontinuity, $\Delta T$, across the boundary between the nickel nanostructures and the substrate, is introduced with the equation:

$$\mathbf{n} \cdot (K_{bulk} \nabla T) = -\frac{\Delta T}{r_{eff}} \quad \text{(S.3)}$$

where $\mathbf{n}$ is the unit vector normal to substrate surface.

The simulations provide a time-dependent surface deformation profile, as illustrated in Supp. Fig. S3b. Fresnel optical propagation is used to calculate the dynamic diffraction signal from these deformations which can be directly compared to our experimental observations. For each sample geometry, we calculate the diffraction signals from a comprehensive set of deformation profiles related to different values of $r_{eff}$. The effective resistivity $r_{eff}$ that then provides the best fit to the experimental data is selected to represent the sum of the constant intrinsic thermal boundary resistivity and corrections due to the nanoscale size effects. All the material properties used in the finite element multiphysics modeling are listed in Supplementary Table S1. Given the maximum temperature change induced in our experiment of $\leq 40$ K, the specific heat and bulk thermal conductivity can change by $\leq 10\%$, and we confirm that any change of this magnitude does not cause appreciable differences in the simulated time-dependent diffraction signal we use to compare with experimental data.

| Material properties | Nickel | Silicon | Sapphire |
|---|---|---|---|
| $C_p$, Specific heat (at 300 K) [J/(kg K)] | 456.8 [1] | 710.0 [24] | 657.5 [1] |
| $K_{bulk}$, Bulk thermal conductivity [W/(m K)] | 90.9 [24] | 149.0 [25] | 41.1 [1] |
| Debye temperature (K) [26] | 450 | 645 | 1047 |
| Poisson's ratio [15] | 0.31 | 0.27 | 0.25 |
| Young's modulus [$10^{11}$ Pa] | 2.00 [27] | 1.31 [28] | 3.45 [29] |
| $\alpha$, Linear coefficient of thermal expansion [$10^{-6}$/K] | 12.77 [1] | 3.00 [30] | 5.21 [1] |
| $\rho$, Density [kg/m$^3$] | 8910 [1] | 2330 [31] | 4000 [1] |

**Supplementary Table S1.** Material parameters used in multiphysics simulations.

## S3. Simple model for interacting heat sources

In this work we introduce a new model to account for the size effects observed in our measurement by using a suppression filter in the phonon MFP spectrum of differential thermal conductivity. The model is derived from the physical limits of diffusive heat transport both for large-linewidth structures and for the case of uniform heating when period equals linewidth, as well as the analytically-derived resistivity correction for small isolated line heat sources[1] and a particular MFP, $\Lambda$:

$$r_{Corr,\ iso} \propto \frac{\Lambda}{(L/2)} \quad (S.4)$$

In addition, we assume that the filter function should be smooth, and that the effects of $L$ and $P$ are uncoupled and the same but opposite to each other. The relevant non-dimensional variables are $L/\Lambda$ and $P/\Lambda$. All of this behavior is encapsulated by a special case of the generic family of logistic functions:

$$S_L(L/\Lambda) = \tanh\left(\frac{L}{2\Lambda}\right) \quad (S.5)$$

$$S_P(P/\Lambda) = 1 - \tanh\left(\frac{P}{2\Lambda}\right) \quad (S.6)$$

$$S_{total}(L, P, \Lambda) = S_L + S_P \quad (S.7)$$

These functions are plotted in Supp. Fig. S4a where we can see the similarity between the shape of $S_{total}$ and standard notch filters. This suppression function can then be applied to each individual MFP-dependent contribution to thermal condutivity to calculate an effective $K_{nano} = \sum_i k(\Lambda_i) \cdot S(L, P, \Lambda_i)$

The resistivity correction for heat transport from interacting nanoscale heat sources can now be related to the change in conductivity imposed by the suppression function:

$$r_{Corr,\ int} \propto \frac{1}{K_{nano}} - \frac{1}{K_{bulk}} \quad (S.8)$$

which gives:

$$r_{Corr,\ int}(L, P) = A \left( \frac{1}{\sum_i k(\Lambda_i) S(L, P, \Lambda_i)} - \frac{1}{\sum_i k(\Lambda_i)} \right) \quad (S.9)$$

The proportionality constant, $A$, incorporates geometrical constants along with the length scale introduced by the transformation from conductivity to resistivity. We associate this length scale with a finite region near the heat source in which effective conductivity is modified from the bulk value, and the present model assumes that this length scale is the same for all phonon modes. However, all our simulations and fitting procedures for analyzing our data exist in the effective resistivity picture so that we never need to assume what this length scale might be.

Using the gray model assuming a single MFP (as in the blue curves found in Fig. 2b), the derived $r_{Corr,\ int}$ yields a reasonably good approximation for sapphire, but not for silicon. Both are well fit by including the extension to multiple MFPs.

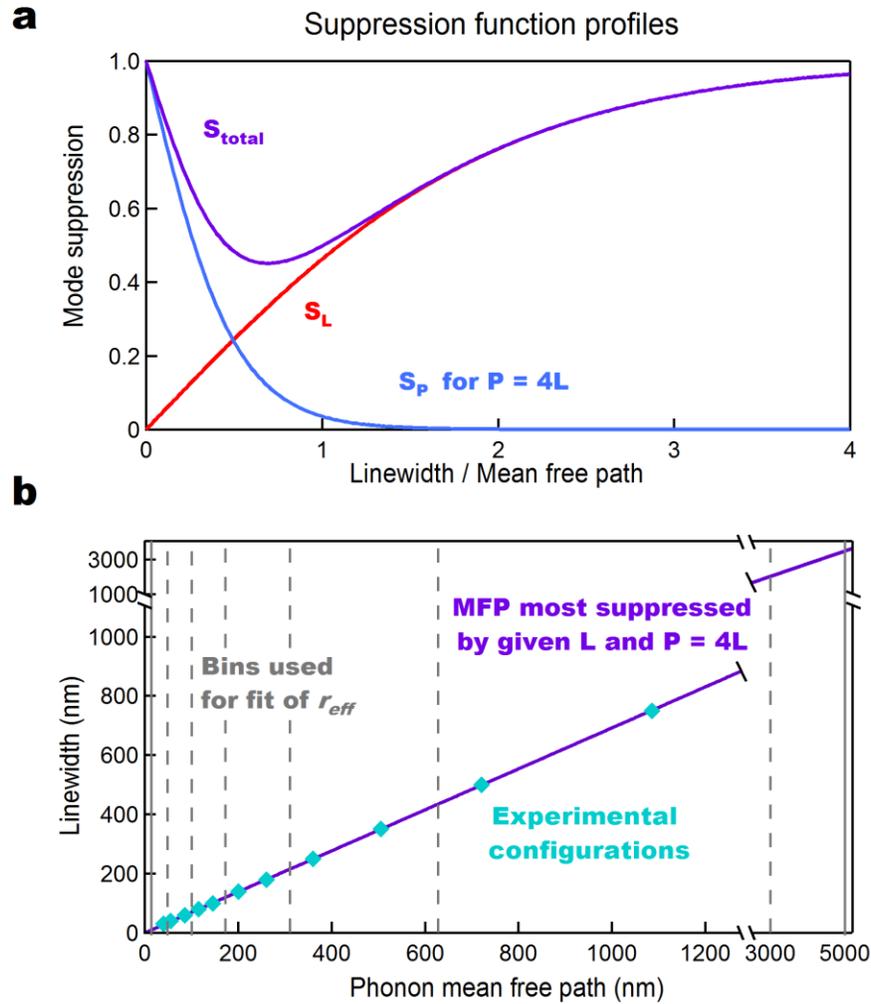

**Supplementary Figure S4 | Visualizing phonon mode suppression. a**, For a phonon mode with a given mean free path, $S_L$ describes the suppression of this mode's contribution to thermal conductivity as the linewidth of a heat source decreases. $S_P$ undoes this suppression, and $S_{total}$ represents the total suppression when both small heat source and interaction between heat sources are taken into account. Thus, each configuration for $L$ and $P$ can be related to one most-suppressed MFP (minimum of $S_{total}$). **b**, We use this information to set the MFP bins used when fitting $r_{eff}$ data in the full interacting multi-MFP model.

Our model can be used in combination with calculated differential conductivity distributions (like the one shown in Fig. 3) to test how well they can account for our observations of $r_{eff}$, including in particular the double-peak structure we observe for the silicon substrate. It can also be inverted to allow the extraction of differential conductivity information from $r_{eff}$ data, as shown in Fig. 4.

The upper and lower bounds of the full range of MFP contributions to which we are sensitive are set by the suppression functions related to each of our nano-gratings. We choose the minimum (14 nm) and maximum (5 μm) MFPs of our experimental spectrum to include only MFPs that are suppressed by at least 20% in our smallest and largest sample geometries, respectively. As can be seen in Supp. Fig. S4a, each particular configuration for *L* and *P* can be related to one most-suppressed MFP – at the minimum of $S_{total}$. We use this information from our set of nano-gratings to set the MFP bins which we use when fitting $r_{eff}$ data in the full interacting multi-MFP model, as shown in Supp. Fig. S4b: bin boundaries are chosen halfway between neighboring most-suppressed MFPs.

The weights $k(\Lambda_i)$ fit to each bin give their relative contribution per nanometer to the differential thermal conductivity. The error bars in the histograms of Fig. 4 are obtained by varying $k(\Lambda_i)$ while maintaining the residual of the fit within the range of experimental uncertainty.

The analytical model described by equation (S.9) also reveals the connection between $r_{eff}$ and the effective thermal conductivity $K_{nano}$ for each linewidth-period pair. This can allow a more direct comparison with those experimental geometries without interfaces, which measure an effective conductivity. Specifically, the relative effective conductivity is given by:

$$\frac{K_{nano}}{K_{bulk}}(L,P) = \frac{A}{K_{bulk}} \left( r_{eff}(L,P) - r_{TBR} + \frac{A}{K_{bulk}} \right)^{-1} \qquad (S.10)$$

It is important to note that different experimental geometries (for example, 1D- versus 2D-confined heat sources, or bulk materials compared with thin films compared with nanotubes) will result in observed conductivities which are specific to the given geometry. Consequently, effective thermal conductivity results cannot necessarily be compared in a straightforward manner. However, the phonon MFP spectrum corresponds to a physically real attribute of a material alone and therefore provides the more appropriate tool for comparison across different experimental geometries. Furthermore, the effective thermal conductivity for any experimental geometry can then be predicted using an experimental phonon MFP spectrum combined with the appropriate theoretical model for conductivity suppression.

## S4. Thermal conductivity and phonon mean free path spectra from first-principles calculations

First principle-based methods for calculating the phononic thermal conductivity for a variety of bulk and nanostructured materials have recently been developed[20,32,33], where great agreement with the measured thermal conductivity values at different temperatures has been demonstrated. These calculations can differentiate the contributions to the total thermal conductivity of phonons with different wavelengths and different mean free paths. Here, we follow the work by Esfarjani et al.[20] to calculate the thermal conductivities and phonon mean free path distributions of silicon and sapphire under the relaxation time approximation.

In this approach, the second-order harmonic and third-order anharmonic interatomic force constants are extracted from first-principles calculations using the direct method[34]. We first record the net forces on all atoms in a supercell ($2\times2\times2$ conventional unit cells for silicon and $3\times3\times3$ primitive unit cells for sapphire) when one or two atoms are displaced from their equilibrium positions by steps of 0.015 Å up to a maximum displacement, $d_{cutoff}$. A fit to the resulting force-displacement curves yields both the harmonic second-order and anharmonic third-order force constants[34]. For the harmonic interaction, $d_{cutoff}$ is set to 5.8 Å for silicon and 5.1 Å for sapphire, while the cutoffs for the third-order anharmonic interaction are 3.9 Å and 2.5 Å, respectively. The cutoffs are chosen by considering the balance between accuracy and computational power. In particular, the cutoffs for silicon are similar to the values used by Esfarjani et al.[20], which reproduced the experimental measurements of silicon phonon dispersion and thermal conductivity very well. Due to the complexity of the crystal structure of sapphire, we have to limit the interaction within a smaller range so that the number of force constants is reasonable to handle. Although we use a relatively small cutoff for sapphire, we impose translational invariance[34] to make sure that the force constants extracted are physically reasonable. All first-principles calculations are performed using the Vienna *ab initio* Simulation Package[35] with the projector augmented wave pseudopotential[36] and the local-density approximation functional. The kinetic-energy cut-off for the plane-wave basis is set at 500 eV. $4\times4\times4$ and $2\times2\times2$ *k*-meshes are used to sample the reciprocal space of silicon and sapphire, respectively. The choice of the energy cutoff and *k*-mesh ensures that the energy change is smaller than 1 meV/atom when refining these two parameters.

After extracting the harmonic force constants, the phonon dispersion relation can be calculated. Supplementary Figure S5 shows the calculated phonon dispersion curves and phonon density of states (DOS) for silicon and sapphire. We also plot the phonon dispersion measured by neutron scattering experiments[37,38] as well as the DOS from other first-principles calculations[20,39]. The good agreement with both the measurements and other theoretical work validates the interatomic force constants from our first-principles calculations.

Using the anharmonic third-order force constants, the scattering rate of three-phonon process can be calculated using Fermi's golden rule[20,40]. The phonon lifetime $\tau_{\mathbf{q}s}$ of each mode $\mathbf{q}s$ (the $s^{th}$ mode at wavevector $\mathbf{q}$) is computed by summing up the scattering rates of all possible three-phonon scattering events.

Defining the mean free path $\Lambda_{\mathbf{q}s}$ as $v_{\mathbf{q}s}\tau_{\mathbf{q}s}$, where $v_{\mathbf{q}s}$ is the group velocity of mode $\mathbf{q}s$, the phonon thermal conductivity can be expressed as a function of $\Lambda_{\mathbf{q}s}$:

$$K = \frac{1}{3}\sum_{\mathbf{q}s} C_{\mathbf{q}s} v_{\mathbf{q}s} \Lambda_{\mathbf{q}s} \quad\quad (S.11)$$

where $C_{\mathbf{q}s}$ is the mode heat capacity. The calculated temperature-dependent thermal conductivity of silicon and sapphire is presented in Supp. Fig. S6, as well as predictions for silicon from other theoretical calculations based on similar approaches[20,41] and experimental data for both silicon[42] and sapphire[43]. Our calculations are in good agreement with these previous works.

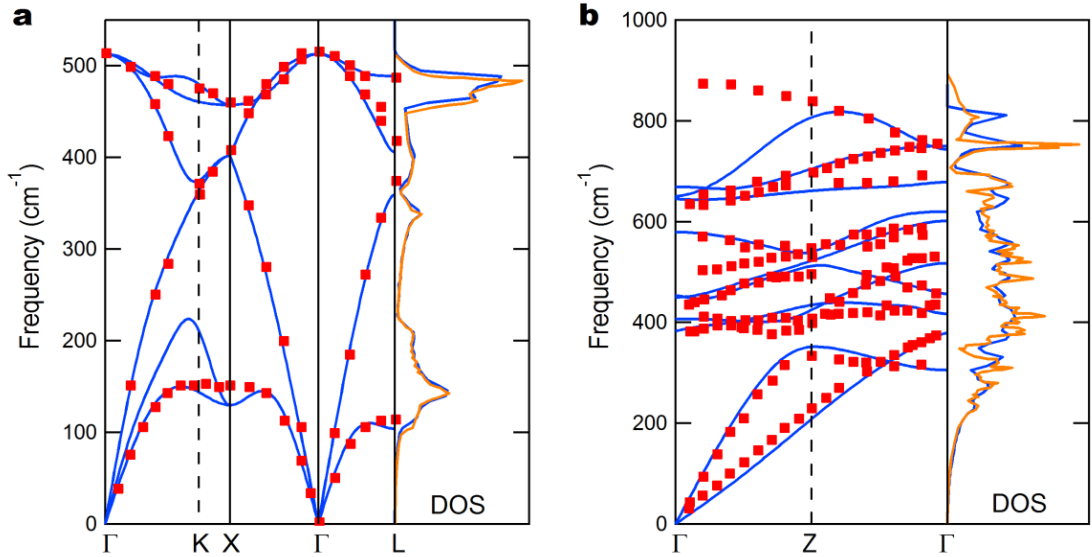

**Supplementary Figure S5 | Comparing our DFT calculations with published work.** Calculated phonon dispersion relation and DOS of silicon **a**, and sapphire **b**, are shown in blue. The red dots in (**a**) represent the measured silicon phonon dispersion by Nilsson and Nelin[37], while the orange curve is the DOS of silicon calculated by Esfarjani et al.[20] The red dots in (**b**) represent the measured sapphire phonon dispersion by Schober et al.[38], and the orange curve is the calculated DOS by Heid et al.[39]

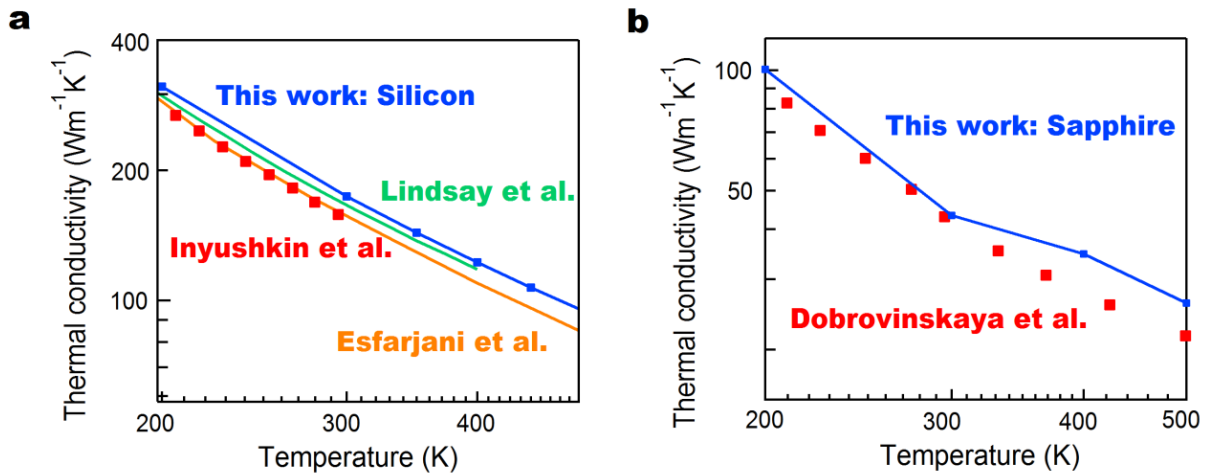

**Supplementary Figure S6 | Temperature-dependent thermal conductivity.** Calculated thermal conductivity curves for silicon **a**, and sapphire **b**, are shown in blue. The red dots in (**a**) represent the measured thermal conductivity of isotope-enriched silicon by Inyushkin et al.[42] while the orange and green curves are other previous theoretical calculations[20,41]. The red dots in (**b**) show the experimentally measured thermal conductivity for sapphire[43].

The differential thermal conductivity $(dK/d\Lambda)\Delta\Lambda$ gives the contribution from the phonons with MFP between $\Lambda - \Delta\Lambda/2$ and $\Lambda + \Delta\Lambda/2$ to the total thermal conductivity. By summing up the thermal conductivity of phonon modes with MFPs in that range, the differential thermal conductivity can be determined. Supplementary Figure S7 shows the differential and cumulative thermal conductivity of silicon from first-principles calculations. Our calculated cumulative thermal conductivity for silicon is similar to the work by Esfarjani et al.[20]

The differential thermal conductivity curve illustrated in Supp. Fig. S7 is discontinuous because only a finite number of sampling points is used in the first Brillouin zone. Since only the MFP of phonon modes at these points can be accurately evaluated, it is a common practice to assume each sampling point represents a neighboring region[20], as illustrated by the red box in Supp. Fig. S8, in which all phonon modes have the same MFP and mode thermal conductivity. As a result, the differential thermal conductivity in Supp. Fig. S7 displays discrete spikes in MFP. In order to obtain the continuous mean free path spectrum, we linearly interpolate the MFP of the phonon modes between the sampling points along the direction of a reciprocal vector, while assuming the mode thermal conductivity is the average value of the neighboring phonon modes on the mesh. This interpolation procedure is illustrated in Supp. Fig. S8 and is found to preserve the total thermal conductivity and the shape of the cumulative thermal conductivity function, as shown in Supp. Fig. S9.

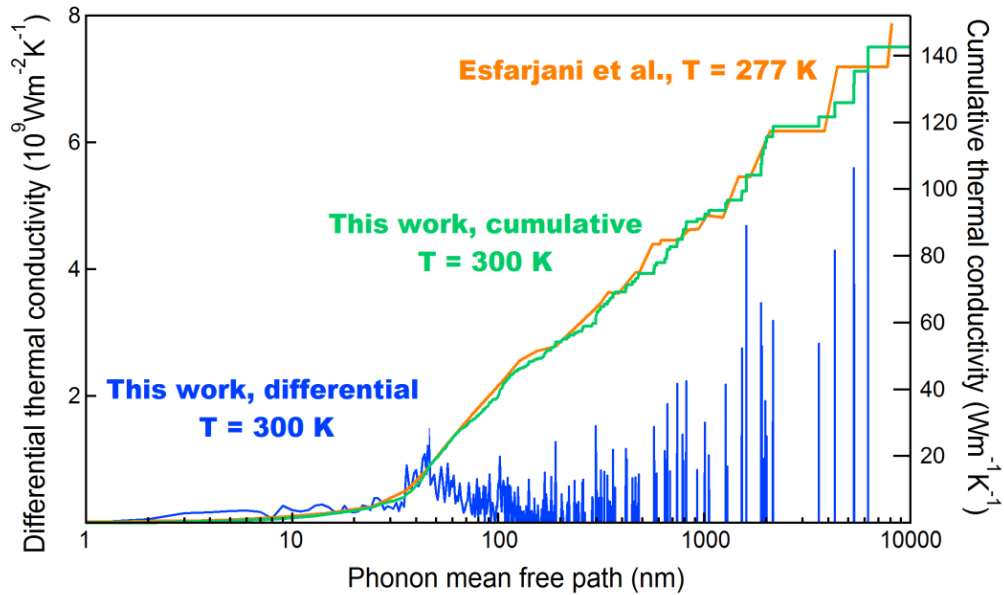

**Supplementary Figure S7 | Thermal conductivity MFP spectra.** The differential (blue) and cumulative thermal conductivity (green) of silicon are reported. We also show the cumulative thermal conductivity calculated by Esfarjani et al.[20] (orange).

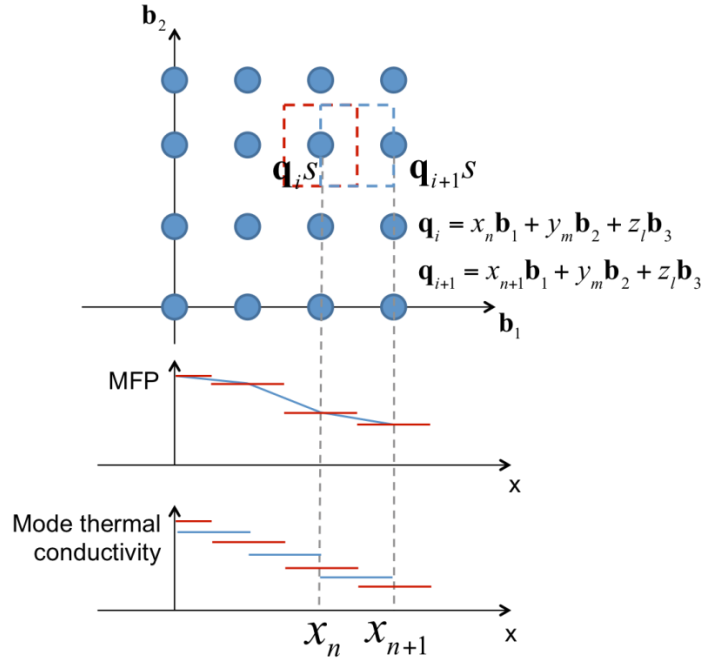

**Supplementary Figure S8 | Schematic of the interpolation process.** The first-principles calculations assume that all phonon modes contained within the red box around mesh-point $q_i s$ have the same MFP and thermal conductivity as the mode on the mesh point, resulting in discrete spikes in the differential conductivity spectrum. To build the continuous function implied by the discrete spectrum, we interpolate the MFP of phonons in the region represented by the blue box in between the neighboring mesh points and assign to them the average mode thermal conductivity as shown in the bottom two graphs.

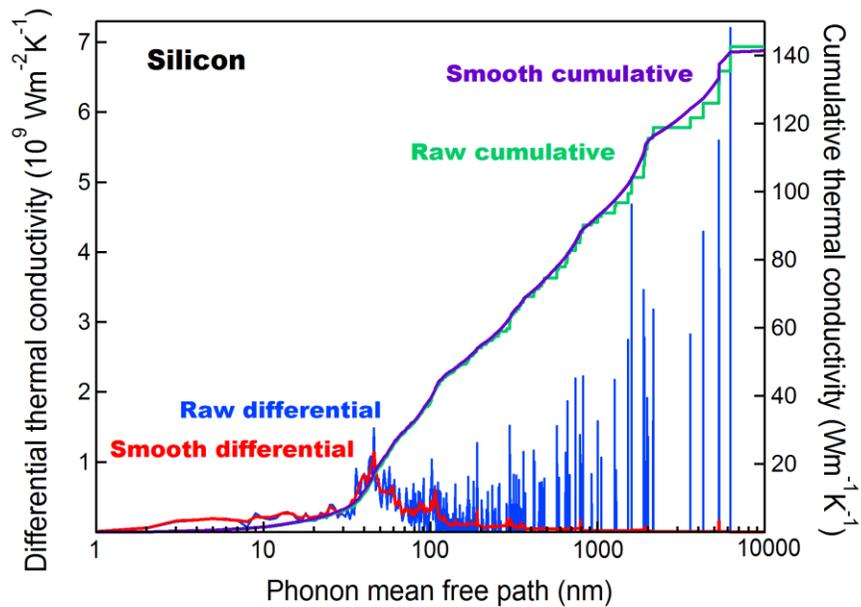

**Supplementary Figure S9 | Interpolating the discrete spectrum.** The smoothed differential (red) and cumulative thermal conductivity (purple) compare well with the discrete, raw spectra (blue and green, respectively).